# A Complete Analysis of Subgiant Stellar Systems III: Hip70868

Hoda E. Elgendy[1(✉)], Mashhoor A. Al-Wardat[1,2,3], Hassan B. Haboubi[1], Lin R. Benchi[1], Abdallah M. Hussein[3], and Hussein M. Elmehdi[1]

[1] Department of Applied Physics and Astronomy, University of Sharjah, Sharjah 27272, UAE
U20104142@sharjah.ac.ae, Hoda.elgendy@ese.gov.ae
[2] Sharjah Academy for Astronomy, Space Sciences and Technology, Sharjah 27272, UAE
[3] Department of Physics, Faculty of Sciences, Al Al-Bayt University, PO Box: 130040, Mafraq, Jordan

**Abstract.** This study utilizes "Al-Wardat's method for analysing binary and multiple stellar systems" to estimate a set of parameters for the binary system Hip70868. The method compares the system's observational magnitudes, color indices, and spectral energy distribution (SED) and synthetic SEDs generated through atmospheric modeling of each component. Feedback-adjusted parameters and an iterative approach were employed to achieve the best fit between observational (including the latest measurements of Gaia, release DR3) and synthetic spectral energy distributions. The findings were completed using the distance of 71.89 pc given by Hipparcos 2007's new reduction. The individual components' parameters for the system were derived afterward. The parameters obtained for the individual components are as follows: Component A has an effective temperature ($T_{eff}$) of $6072 \pm 50$ K, a surface gravity ($\log g_a$) of $4.50 \pm 0.05$, and a radius ($R_a$) of $1.40 \pm 0.07$ $R_\odot$. Component B has a ($T_{eff}$) of $5887 \pm 50$ K, a ($\log g_b$) of $4.40 \pm 0.05$, and a radius ($R_b$) of $1.326 \pm 0.07$ $R_\odot$. The spectral types of the components were found to be F9 IV and G1 IV, respectively. The findings from this analysis were utilized to accurately position the two system components on the Hertzsprung-Russell (H-R) diagram and the evolutionary tracks. The analysis shows that the components have transitioned from the main sequence to the sub-giant stage of their evolution.

## 1 Introduction

Numerous studies indicate that over half the stars in our solar region and galaxy belong to the binary or multiple-star group systems [1]. This fascinating discovery highlights the structures and levels of organization that can be observed within the cosmos. The way these stellar systems form and evolve their architecture is closely linked to factors that include their period ratios, orbit orientation, and mass distribution [2]. This information provides insights into the process of star formation in general. Understanding the characteristics of higher-order systems is significant as most stars are born within groups sharing similar properties [3]. The presence of companions has impacts on the life and eventual death of stars [4]. Thus, understanding the formation, evolution, and behavior of stellar groups requires a thorough analysis of star systems and their binary features. This is widely backed by scientific research papers [5].

The physical and geometrical parameters of binary and multiple star systems are analyzed using "Al-Wardat's method for analyzing binary and multiple stellar systems." This method blends atmospheric modeling with spectrophotometry. The geometrical and physical properties of the components of stars are predicted by this method for several stars (for example see [6–12]). It uses the estimated metallicities and ages of binary and multiple stellar systems (BMSSs) to help explain the formation process [4]. The technique applies to all kinds of binaries, including eclipsing, and visual ones that have not yet reached the giant stage [8, 13].

To apply this method, one must determine the color indices and visual magnitude in addition to the magnitude differences (obtained via measurements made using speckle interferometry). Numerous binary systems have used this method since 2002 [10].

The two components that make up the Hip 70868 system, A and B, are presented in this document alongside their geometrical and physical specifications. The selection of this system was based on its representation of a sub-giant star system, about which there appears to be little information because it is in between the main system and the red giant stages of growth [14]. Studying sub-giant systems is crucial to understanding the transition from main sequence to red giant stages. Further, it helps clarify the unique characteristics of stars in this "intermediate" level [12].

**Table 1.** Basic data of the system Hip 70868.

| Properties | Parameters | Value | Ref |
|---|---|---|---|
| Position | $\alpha_{2000}$ | $14^h\ 29^m\ 32^s.61$ | (*SIMBAD*) |
|  | $\delta_{2000}$ | $-37°\ 02'\ 22''.33$ |  |
| Magnitude [mag] | $m_v$ | 7.84 | [15] |
|  | $(B-V)_J$ | $0.626 \pm 0.015$ | [15] |
|  | AV | $0.0248 \pm 0.002$ | [16] |
| Parallax [mas] | $\pi_{Hip1}$ | $14.79 \pm 1.10$ | [15] |
|  | $\pi_{Hip2}$ | $13.91 \pm 0.91$ | [17] |
|  | $\pi_{GDR2}$ | $12.5757 \pm 0.5952$ | [18] |

## 1.1 Hip 70868

Hip 70868, or HD 126935, belongs to the G3V spectral classification type in the Henry Draper Catalog [19]. The components of binary stars are distinguished by the suffixes "A" (for the brighter or primary component) and "B" (for the faint secondary component in the system). The fundamental parameters of hip 70868 are presented in Table 1.

The system's parallax comes in three different values. The first parallax, with a value of 14.79 mas and a distance of 67.61 pc, is derived from literature catalogs [15]. The second value, 13.91 mas, equivalent to a distance of 71.89 pc, is taken from the Hipparcos reduced catalog [17]. The third value, 12.5757 mas, corresponds to a distance of 78.38 pc and was taken from the Gaia Release 2 [18]. In this paper, the parallax of the reduced Hipparcos catalog is used for the analysis to calculate the distance of the system.

## 2 Atmospheric Modeling

The components of hip 70868's atmospheric properties were identified by applying "Al-Wardat's method for analyzing binary and multiple stellar systems". Using grids of line-blanketed model atmospheres ATLAS 9, this method focuses on modeling entire synthetic spectral energy distributions (SEDs) using preliminary parameters. The "Al-Wardat" atmospheric modeling approach requires additional data to analyze the components of the hip 70868 system.

The magnitude difference between the components $\Delta m = 0.323$ (from *Fourth Catalog*) is used as the average of all measurements done with visual filters $550 \pm 90$ nm in Eq. 1, as well as the visual magnitude $m_v = 7.84$ from Table 1. We employed Eq. 2 to get the preliminary magnitudes for each component. The magnitudes of the primary and secondary components of the system were found to be $m_v^A = 8.46$ mag and $m_v^B = 8.74$ mag respectively.

$$\frac{f1}{f2} = 2.512^{-\Delta m} \quad (1)$$

$$m_v = -2.5 \log(f1 + f2) \quad (2)$$

Additionally, we can compute the initial input parameters for each component, including bolometric magnitudes, luminosities, and effective temperatures using Eq. 3, Eq. 4, and Eq. 5.

$$M_v = m_v + 5 - 5\log(d) - A_v \quad (3)$$

$$\log\left(\frac{R}{R_\odot}\right) = 0.5 \log\left(\frac{L}{L_\odot}\right) - 2\log\left(\frac{T}{T_\odot}\right) \quad (4)$$

$$\log(g) = \log\left(\frac{M}{M_\odot}\right) - 2\log\left(\frac{R}{R_\odot}\right) + 4.43 \quad (5)$$

The preliminary input parameters were calculated with $T_\odot = 5777$ K, of $R_\odot = 6.69 \times 10^8$ m, $A_v = 0.2268$, the dynamic parallax $13.91 \pm 0.91$ (d = 71.89 pc) from Table 1 (to calculate the magnitude) and finally, the bolometric correction from [20].

Employing solar metallicity model atmospheres as the ATLAS 9 initial parameter output, the total luminosities of components A and B, which are located at a distance d (pc) from Earth, combine to form the total energy flux from the system as follows [6, 21]:

$$F_\lambda \cdot d^2 = H_\lambda^A R_A^2 + H_\lambda^B R_B^2 \quad (6)$$

Rearranging

$$F_\lambda = \left(\frac{R_A^2}{d^2}\right)^2 \left[H_\lambda^A + H_\lambda^B \cdot \left(\frac{R_B}{R_A}\right)^2\right] \quad (7)$$

where $H_\lambda^A$ and $H_\lambda^B$ are fluxes from the unit surface of each component. $F_\lambda$ is the total spectral energy density of the system. $R_A$ and $R_b$ are the radii of primary and secondary components in solar units. "Al-Wardat's method" was used to create their synthesized Spectral Energy Distributions to calculate the binary system's magnitudes and color indices.

This technique was also applied as a calibration technique, whereby the synthetic magnitudes of a whole stellar system were compared with available observational data. The precision of the observations determines how accurate the results are.

## 3 Results and Conclusion

The resulting whole synthetic SED is not going to fit the first trial's observational spectral energy density. To find the best fit between the synthetic and observational total absolute fluxes, numerous attempts were conducted. This resulted in the construction of hundreds of synthetic SEDs using various parameter settings and comparisons with the observational SED. Efforts were made to alter the radii, parallax, effective temperatures, and surface gravity acceleration. As our objective preliminary parameters, we consider the values of the total observational $V_J$, $B_T$, $V_T$, and $\Delta m$. To sum up, the parameters were selected for the best fit (Table 2).

**Table 2.** Parameters of the system Hip70868 components as estimated using Al-Wardat's Method.

| Physical Parameter | Value |
| --- | --- |
| $T_A$ | $6072 \pm 50$ K |
| $T_B$ | $5887 \pm 50$ K |
| $\log g_A$ | $4.50 \pm 0.05$ |
| $\log g_B$ | $4.40 \pm 0.05$ |
| $R_A$ | $1.40 \pm 0.07$ $R_\odot$ |
| $R_B$ | $1.326 \pm 0.07$ $R_\odot$ |
| d | 71.89 pc |

The synthetic data for each component and the entire data for the system are shown in Table 3. The spectral energy distributions obtained by the synthetic analysis are shown in Fig. 1. According to the tables in [20], the spectral types of the components are F9-IV and G1-IV, respectively.

The luminosity of each component is calculated using Eq. 4 and the error of luminosity values is calculated through the following method:

$$\sigma_L = \pm 2L \sqrt{\left(\frac{\sigma_R}{R}\right)^2 + 4\left(\frac{\sigma_{Teff}}{Teff}\right)^2} \quad (8)$$

**Table 3.** Magnitudes and color indices of the composed synthetic spectrum and individual components of Hip70868 as a result of applying the method. The entire values were used for the best fit.

| Sys | Filter | Entire Obs | Entire Synt | HIP70868 | |
|---|---|---|---|---|---|
| | | | $\sigma = \pm 0.03$ | A | B |
| Joh- | U | 8.48 | 8.60 | 9.17 | 9.57 |
| Cou | B | | 8.47 | 9.06 | 9.34 |
| | V | 7.84 | 7.84 | 8.46 | 8.74 |
| | R | $0.626 \pm 0.015$ | 7.50 | 8.13 | 8.39 |
| | U – B | | 0.13 | 0.10 | 0.17 |
| | B – V | | 0.63 | 0.60 | 0.66 |
| | V – R | | 0.34 | 0.33 | 0.35 |
| Stromgren | u | | 9.74 | 10.32 | 10.72 |
| | v | | 8.80 | 9.39 | 9.75 |
| | b | | 8.19 | 8.80 | 9.11 |
| | y | | 7.81 | 8.43 | 8.71 |
| | u – v | | 0.94 | 0.92 | 0.97 |
| | v – b | | 0.61 | 0.59 | 0.64 |
| | b – y | | 0.38 | 0.37 | 0.397 |
| Tycho | $B_T$ | $8.631 \pm 0.010$ | 8.62 | 9.20 | 9.56 |
| | $V_T$ | $7.912 \pm 0.008$ | 7.91 | 8.53 | 8.82 |
| | $B_T - V_T$ | 0.719 | 0.71 | 0.68 | 0.75 |
| Gaia | G | 7.722 | 7.72 | 8.35 | 8.61 |
| | Bp | 8.001 | 8.07 | 8.68 | 8.98 |
| | Rp | 7.211 | 7.20 | 7.83 | 8.07 |
| Infrared | J | 6.709 | 6.71 | 7.37 | 7.56 |
| | H | 6.479 | 6.42 | 7.09 | 7.25 |
| | K | 6.361 | 6.38 | 7.06 | 7.22 |

Yielding $L_A = (2.39 \pm 0.25)$ $L_\odot$ and $L_B = (1.896 \pm 0.210)$ $L_\odot$ luminosities for each component. This shows that our system has begun evolving into the sub-giant stage according to the evolutionary tracks shown in Fig. 2 [22].

Overall, we estimated the atmospheric parameters of each component of HIP70868 using "Al-Wardat's method for analyzing BAMSs". Using the estimated parameters, "Al-Wardat's method" builds a synthetic spectral energy distribution for each component (SED) and provides the position of the components' system on the H-R diagram with the evolutionary tracks. The parameters of the components are shown in Table 4.

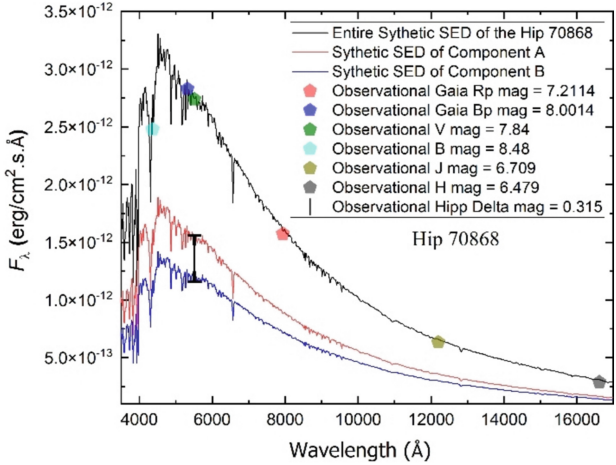

**Fig. 1.** Synthetic Spectral Energy Distribution of the entire Hip 70868 system and its individual components. The complete synthetic SED aligns well with the observed magnitudes from various sources, including the latest Gaia DR3 data.

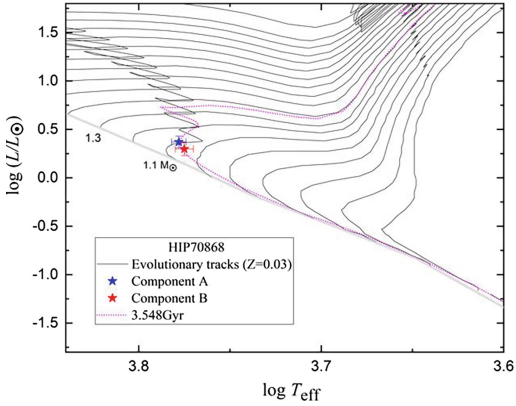

**Fig. 2.** The components of Hip 70868 on the evolutionary tracks and isochrons of (Girardi et al., 2000).

**Table 4.** Parameters of the individual components of Hip70868 (This work).

| Parameter | Unit | Component A | Component B |
|---|---|---|---|
| $T_{eff}$ | K | $6072 \pm 50$ | $5887 \pm 50$ |
| logg | $cm/s^2$ | $4.50 \pm 0.05$ | $4.40 \pm 0.05$ |

(*continued*)

Table 4. (*continued*)

| Parameter | Unit | Component A | Component B |
|---|---|---|---|
| R | $R_\odot$ | 1.40 ± 0.07 | 1.326 ± 0.07 |
| L | $L_\odot$ | 2.39 ± 0.25 | 1.89 ± 0.21 |
| $M_{bol}$ | Mag | 3.80 | 4.05 |
| M | $M_\odot$ | 1.22 | 1.16 |
| d | pc | 71.89 | |
| Age | Gyr | 3.548 | |

## 4  Orbital Analysis

Using the IDL code ORBITX developed by Tokovinin [23, 24], and inputting the orbital data shown in Tables 5 and 6, we get an improved orbit of our system shown in Fig. 3.

Table 5. New orbital elements of Hip70868.

| Parameters | Units | Values |
|---|---|---|
| P [Period] | [Yr] | 14.18830 ± 0.01593 |
| T [Epoch of passage] | | 2008.6061 ± 0.0024 |
| e [Eccentricity] | - | 0.4398 ± 0.0012 |
| a [Semi-major] | [Arcsec] | 0.1173 ± 0.0002 |
| ω [Longitude] | [deg] | 167.10 ± 0.16 |
| Ω [Position angle] | | 125.17 ± 0.27 |
| i [Inclination] | | 128.15 ± 0.18 |

According to Fig. 3 we notice that all 12 points are aligned in the orbit starting from 1991 to the 2021 observations. Using this data, we can use Eq. 9 and Eq. 10 to find the mass and error of the system.

$$\text{Mdyn} = \text{MA} + \text{MB} = \left(\frac{a^3}{\pi^3 P^2}\right) M_\odot \qquad (9)$$

$$\frac{\sigma_M}{M} = \sqrt{9\left(\frac{\sigma_\pi}{\pi}\right)^2 + 9\left(\frac{\sigma_a}{a}\right)^2 + 4\left(\frac{\sigma_p}{p}\right)^2} \qquad (10)$$

Yeilding a total dynamical mass of (2.48 ± 0.55) $M_\odot$ using parallax $\pi_{\text{Hip1}} = 14.79$ mas and a total dynamical mass of (2.98 ± 0.58) $M_\odot$ using parallax $\pi_{\text{Hip2}} = 13.91$ mas from Table 1.

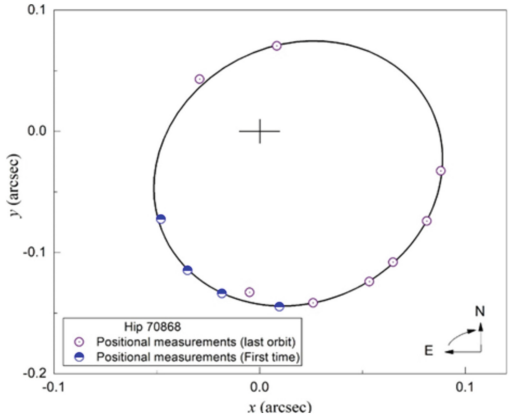

**Fig. 3.** New orbit of Hip 70868.

**Table 6.** The relative positional measurements of the system as of the Fourth Catalog of Interferometric Measurements of Binary Stars.

| Date | θ (deg) | ρ (deg) |
| --- | --- | --- |
| 1991.250 | 178.0 ± 0.133 | 0.133 |
| 2008.542 | 34.1 ± 0.052 | 0.052 |
| 2009.260 | 353.4 ± 0.071 | 0.071 |
| 2013.132 | 249.8 ± 0.094 | 0.094 |
| 2014.303 | 227.8 ± 0.110 | 0.11 |
| 2015.497 | 211.1 ± 0.126 | 0.126 |
| 2016.134 | 203.4 ± 0.135 | 0.135 |
| 2017.430 | 190.5 ± 0.144 | 0.144 |
| 2018.162 | 183.9 ± 0.145 | 0.145 |
| 2019.372 | 172.3 ± 0.135 | 0.135 |
| 2020.117 | 163.1 ± 0.120 | 0.12 |
| 2021.078 | 146.6 ± 0.087 | 0.087 |

## 5  Conclusion

The method of analyzing binary or multiple stellar systems, which was developed by Mashhoor Al- Wardat, and the orbital analysis method, which was developed by Adrei A. Tokovinin, were applied to the visually close binary star HIP 70868. Based on the best fit between the system's synthetic SED and observational photometry, the atmospheric, fundamental, and orbital parameters have been computed. The results indicated a 3.548 Gyr old subgiant star, with F7 IV component A and a G1 IV component B. Furthermore, the system's total and partial synthetic magnitudes and color indices were calculated using the Hipparcos new reduction parallax of 13.91 mas. Finally, using orbital data, the dynamical mass of the system was calculated as well.

**Acknowledgement.** This work has made use of data from the European Space Agency (ESA) mission Gaia (https://www.cosmos.esa.int/gaia), processed by the Gaia Data Processing and Analysis Consortium (DPAC; https://www.cosmos.esa.int/web/gaia/dpac/consortium). This work has made use of the Multiple Star Catalog, SIMBAD database, and Al-Wardat's method for analyzing binary and multiple stellar systems with their codes.